\documentclass[aps,prl,twocolumn,tightenlines,superscriptaddress,showpacs,byrevtex]{revtex4}

\usepackage{epsfig,floatflt}
\usepackage{amssymb,amsmath,citesort,indentfirst,fancyhdr,enumerate}
\usepackage{rotating}
\usepackage{dcolumn}  

\graphicspath{{ps}}

\newcommand{\BF}{\ensuremath{{\cal{B}}}}

\newcommand{\ACP}{\ensuremath{{A_{CP}}}}
\newcommand{\UFS}{\ensuremath{\Upsilon(4S)}}
\newcommand{\bbbar}{\ensuremath{B\bar{B}}}
\newcommand{\qqbar}{\ensuremath{q\bar{q}}}

\newcommand{\de}{\ensuremath{\Delta E}}
\newcommand{\mb}{\ensuremath{M_{\rm bc}}}

\newcommand{\bckpp}{\ensuremath{B^\pm\to K^\pm\pi^\pm\pi^\mp}}

\newcommand{\kpp}{\ensuremath{K^\pm\pi^\pm\pi^\mp}}
\newcommand{\kppp}{\ensuremath{K^+\pi^+\pi^-}}
\newcommand{\kmpp}{\ensuremath{K^-\pi^-\pi^+}}

\newcommand{\pipi}{\ensuremath{\pi^+\pi^-}}
\newcommand{\kcpi}{\ensuremath{K^+\pi^-}}

\newcommand{\chic}{\ensuremath{\chi_{c0}}}

\def\nima#1#2#3{{Nucl.\, Instr.\, and Meth.} {A \bf #1}, #3 (#2)}

\def\npb#1#2#3{{ Nucl.\, Phys.}             {B \bf #1}, #3 (#2)}
\def\plb#1#2#3{{ Phys.\, Lett.}             {B \bf #1}, #3 (#2)}

\def\prd#1#2#3{{ Phys.\, Rev.}              {D \bf #1}, #3 (#2)}
\def\pr#1#2#3 {{ Phys.\, Rev.}              {  \bf #1}, #3 (#2)}

\def\prl#1#2#3{{ Phys.\, Rev.\, Lett.}       {  \bf #1}, #3 (#2)}

\begin{document}

\title{ \boldmath Evidence for Large Direct $CP$ Violation
in $B^\pm\to\rho(770)^0K^\pm$ from Analysis of the Three-Body
Charmless $\bckpp$ Decay}

\affiliation{Aomori University, Aomori}
\affiliation{Budker Institute of Nuclear Physics, Novosibirsk}
\affiliation{Chiba University, Chiba}
\affiliation{Chonnam National University, Kwangju}
\affiliation{University of Cincinnati, Cincinnati, Ohio 45221}
\affiliation{University of Frankfurt, Frankfurt}
\affiliation{Gyeongsang National University, Chinju}
\affiliation{University of Hawaii, Honolulu, Hawaii 96822}
\affiliation{High Energy Accelerator Research Organization (KEK), Tsukuba}
\affiliation{Hiroshima Institute of Technology, Hiroshima}
\affiliation{Institute of High Energy Physics, Chinese Academy of Sciences, Beijing}
\affiliation{Institute of High Energy Physics, Protvino}
\affiliation{Institute of High Energy Physics, Vienna}
\affiliation{Institute for Theoretical and Experimental Physics, Moscow}
\affiliation{J. Stefan Institute, Ljubljana}
\affiliation{Kanagawa University, Yokohama}
\affiliation{Korea University, Seoul}
\affiliation{Kyoto University, Kyoto}
\affiliation{Kyungpook National University, Taegu}
\affiliation{Swiss Federal Institute of Technology of Lausanne, EPFL, Lausanne}
\affiliation{University of Ljubljana, Ljubljana}
\affiliation{University of Maribor, Maribor}
\affiliation{University of Melbourne, Victoria}
\affiliation{Nagoya University, Nagoya}
\affiliation{Nara Women's University, Nara}
\affiliation{National Central University, Chung-li}
\affiliation{National United University, Miao Li}
\affiliation{Department of Physics, National Taiwan University, Taipei}
\affiliation{H. Niewodniczanski Institute of Nuclear Physics, Krakow}
\affiliation{Nippon Dental University, Niigata}
\affiliation{Niigata University, Niigata}
\affiliation{Nova Gorica Polytechnic, Nova Gorica}
\affiliation{Osaka City University, Osaka}
\affiliation{Osaka University, Osaka}
\affiliation{Panjab University, Chandigarh}
\affiliation{Peking University, Beijing}
\affiliation{University of Pittsburgh, Pittsburgh, Pennsylvania 15260}
\affiliation{Princeton University, Princeton, New Jersey 08544}
\affiliation{RIKEN BNL Research Center, Upton, New York 11973}
\affiliation{Saga University, Saga}
\affiliation{University of Science and Technology of China, Hefei}
\affiliation{Seoul National University, Seoul}
\affiliation{Shinshu University, Nagano}
\affiliation{Sungkyunkwan University, Suwon}
\affiliation{University of Sydney, Sydney NSW}
\affiliation{Tata Institute of Fundamental Research, Bombay}
\affiliation{Toho University, Funabashi}
\affiliation{Tohoku Gakuin University, Tagajo}
\affiliation{Tohoku University, Sendai}
\affiliation{Department of Physics, University of Tokyo, Tokyo}
\affiliation{Tokyo Institute of Technology, Tokyo}
\affiliation{Tokyo Metropolitan University, Tokyo}
\affiliation{Tokyo University of Agriculture and Technology, Tokyo}
\affiliation{Toyama National College of Maritime Technology, Toyama}
\affiliation{University of Tsukuba, Tsukuba}
\affiliation{Virginia Polytechnic Institute and State University, Blacksburg, Virginia 24061}
\affiliation{Yonsei University, Seoul}
  \author{A.~Garmash}\affiliation{Princeton University, Princeton, New Jersey 08544} 
  \author{K.~Abe}\affiliation{High Energy Accelerator Research Organization (KEK), Tsukuba} 
  \author{K.~Abe}\affiliation{Tohoku Gakuin University, Tagajo} 
  \author{I.~Adachi}\affiliation{High Energy Accelerator Research Organization (KEK), Tsukuba} 
  \author{H.~Aihara}\affiliation{Department of Physics, University of Tokyo, Tokyo} 
  \author{Y.~Asano}\affiliation{University of Tsukuba, Tsukuba} 
  \author{T.~Aushev}\affiliation{Institute for Theoretical and Experimental Physics, Moscow} 
  \author{T.~Aziz}\affiliation{Tata Institute of Fundamental Research, Bombay} 
  \author{S.~Bahinipati}\affiliation{University of Cincinnati, Cincinnati, Ohio 45221} 
  \author{A.~M.~Bakich}\affiliation{University of Sydney, Sydney NSW} 
% ...
\collaboration{The Belle Collaboration}


\noaffiliation

\begin{abstract}
We report results on a Dalitz analysis of three-body charmless $\bckpp$ decay
including searches for direct $CP$ violation. We report the first observation
of the decay $B^\pm\to f_2(1270)K^\pm$ with a statistical significance above
$6\sigma$. We also observe $3.9\sigma$ evidence for large direct $CP$ violation
in the $B^\pm\to\rho(770)^0K^\pm$ channel. This is the first evidence for $CP$
violation in a charged meson decay. The results are obtained with a data sample
that contains 386 million $B\bar{B}$ pairs collected at the $\Upsilon(4S)$
resonance with the Belle detector at the KEKB asymmetric-energy $e^+e^-$
collider.
\end{abstract}

\maketitle

{\renewcommand{\thefootnote}{\fnsymbol{footnote}}}
\setcounter{footnote}{0}

\normalsize


Decays of $B$ mesons to three-body charmless hadronic final states provide new
possibilities for $CP$ violation searches. In decays to two-body final states
($B\to K\pi$, $\pi\pi$, etc.) direct $CP$ violation can be observed as a
difference in $B$ and $\bar{B}$ decay rates. In decays to three-body final
states that are often dominated by quasi-two-body channels, direct $CP$
violation can also manifest itself as a difference in relative phase between
two quasi-two-body amplitudes that can be measured via amplitude (Dalitz)
analysis. So far direct $CP$ violation has been observed only in decays of
neutral $K$ mesons\,\cite{dcpv-K0} and recently in neutral $B$ meson
decays\,\cite{dcpv-B0}. However, large direct $CP$ violation is expected in
charged $B$ decays to some quasi-two-body charmless hadronic
modes\,\cite{beneke-neubert}.

The search for direct $CP$ violation in the three-body charmless
$B^\pm$$\to$$\kpp$ decay described in this Letter is performed by applying a
Dalitz analysis technique\,\cite{Dalitz} to a data sample containing
386 million $B\bar{B}$ pairs, collected with the Belle
detector\,\cite{Belle} operating at the KEKB asymmetric-energy $e^+e^-$
collider\,\cite{KEKB} with a center-of-mass (c.m.)\ energy at the $\UFS$
resonance. These results supersede the results reported in
Ref.\,\cite{khh-dalitz-belle}.

Charged tracks are required to have momenta transverse to the beam greater than
0.1~GeV/$c$ and to be consistent with originating from the interaction region.
For charged kaon identification we impose a requirement on a particle
identification variable which has 86\% efficiency and a 7\% fake rate from
misidentified pions as measured from data. Charged tracks that are positively
identified as electrons or protons are excluded. $B$ candidates are identified
using two kinematic variables: the energy difference
$\de = (\sum_i\sqrt{c^2|\mathbf{p}_i|^2 + c^4m_i^2} ) - E^*_{\rm beam},$
and the beam constrained mass
$\mb =  \frac{1}{c^2}\sqrt{E^{*2}_{\rm beam}-c^2|\sum_i\mathbf{p}_i|^2},$
where the summation is over all particles from a $B$ candidate; ${\bf p}_i$
and $m_i$ are their c.m.\ three-momenta and masses, respectively;
$E^*_{\rm beam}$ is the beam energy in the c.m.\, frame.
The signal $\mb$ resolution is mainly given by the beam energy spread,
and amounts to 2.9~MeV/$c^2$. The signal $\de$ shape is
fitted by a sum of two Gaussian functions with a common mean. In fits to the
experimental data, we fix the width (35~MeV) and the relative fraction (0.16)
of the second Gaussian function to the values obtained from Monte Carlo (MC)
simulation. The common mean of the two Gaussian functions and the width of
the main Gaussian are allowed to float.

The dominant background is due to $e^+e^-$$\to\,$$\qqbar$ ($q=u,d,s$ and $c$
quarks) continuum events. We reject about 98\% of this background while
retaining 36\% of the signal using variables that characterize the event
topology. For more details see Ref.\,\cite{khh-belle} and references therein.
From MC studies we find that the dominant backgrounds originating from other
$B$ decays that peak in the signal region are due to
$B^+$$\to\,$$\bar{D}^0[K^+\pi^-]\pi^+$ and due to
$B^+$$\to\,$$J/\psi(\psi(2S))[\mu^+\mu^-]K^+$ in which muons are
misidentified as pions. We veto these backgrounds by applying requirements on
the invariant masses of the appropriate two-particle
combinations\,\cite{khh-dalitz-belle}. The most significant backgrounds from
charmless $B$ decays originate from $B^+$$\to$\,$\eta'[\gamma\pipi]K^+$,
$B^+$$\to$\,$\pi^+\pi^+\pi^-$ where one of the two same-charge pions is
misidentified as a kaon, and from $B^0$$\to$$K^+\pi^-$ processes.

\begin{figure}[!t]
\includegraphics[width=0.475\textwidth]{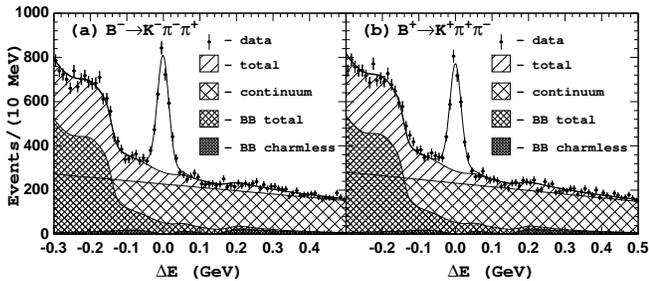}
 \caption{$\de$ distributions for (a) $B^-$$\to$$\kmpp$ and\
          (b) $B^+$$\to$$\kppp$
          events with $|\mb$$-$$M_B|$$<$$7.5$~MeV/$c^2$.
          Points with error bars are data; the smooth curve is the fit 
          result; the hatched areas are various background components.}
  \label{fig:khh-DE}
\end{figure}

The $\de$ distributions for $\bckpp$ candidates that pass all the selection
requirements are shown in Fig.~\ref{fig:khh-DE}. In the fit to the $\de$
distribution we fix the shape of the $\bbbar$ background component from MC
and let the normalization float. The shape of the $\qqbar$ background is
parametrized by a linear function with the slope and normalization as free
parameters of the fit. From these fits we find $2248\pm79$ ($2038\pm76$)
$B^-$ ($B^+$) signal events; the width of the main signal Gaussian is
$15.3\pm0.5$~MeV.

For the amplitude analysis we select events from the $B$ signal region
defined as an ellipse around the $\mb$ and $\de$ mean values:
$\left[\frac{\mb-M_B}{7.5\ {\rm MeV}/c^2}\right]^2$$+$$\left[\frac{\de}{40\ {\rm MeV}}\right]^2$$<$$1$.
The total number of events in the signal region is 7757; the relative fraction
of signal events is $0.512\pm0.012$. The distribution of background events
is determined from analysis of events in the $\mb$$-$$\de$ sideband region.

\begin{table*}[!t]
\caption{Results of the best fit to $\kpp$ events in the $B$ signal region.
The first quoted error is statistical and the second is the model dependent
uncertainty. The quoted $CP$ asymmetry significance is statistical only.}
\medskip
\label{tab:kpp-fit-res}
\centering
  \begin{tabular}{lccccc}
\hline \hline
Channel & Fraction (\%) 
        & $\delta$ ($^\circ$) 
        & $b$ 
        & $\varphi$ ($^\circ$)
        & Asymmetry significance ($\sigma$)\\
\hline \hline
$K^*(892)\pi^\pm$  & $13.0\pm0.8^{+0.5}_{-0.7}$
                     & $0$ (fixed)
                     & $0.078^{+0.040+0.012}_{-0.031-0.003}$
                     & $ -18\pm44^{+5}_{-13}$
                     & $2.6$ \\
$K^*_0(1430)\pi^\pm$ & $65.5\pm1.5^{+2.2}_{-3.9}$
                     & $55\pm4^{+1}_{-5}$
                     & $0.069^{+0.032+0.010}_{-0.030-0.008}$
                     & $-123\pm16^{+4}_{-5}$
                     & $2.7$ \\
$\rho(770)^0K^\pm$   & $ 7.85\pm0.93^{+0.64}_{-0.59}$
                     & $-21\pm14^{+14}_{-19}$
                     & $0.28^{+0.12+0.07}_{-0.09-0.09}$
                     & $-125\pm32^{+10}_{-85}$
                     & $3.9$ \\
$\omega(782)K^\pm$   & $0.15\pm0.12^{+0.03}_{-0.02}$
                     & $100\pm31^{+38}_{-21}$
                     & $0$ (fixed)   & $-$            & $-$   \\
$f_0(980)K^\pm$      & $17.7\pm1.6^{+1.1}_{-3.3}$
                     & $67\pm11^{+10}_{-11}$
                     & $0.30\pm0.19^{+0.05}_{-0.10}$
                     & $-82\pm8^{+2}_{-2}$
                     & $1.6$ \\
$f_2(1270)K^\pm$     & $ 1.52\pm0.35^{+0.22}_{-0.37}$
                     & $140\pm11^{+18}_{-7}$
                     & $0.37^{+0.19+0.11}_{-0.16-0.04}$
                     & $-24\pm29^{+14}_{-20}$
                     & $2.7$ \\
$f_X(1300)K^\pm$     & $ 4.14\pm0.81^{+0.31}_{-0.30}$
                     & $-141\pm10^{+8}_{-9}$
                     & $0.12\pm0.17^{+0.04}_{-0.07}$
                     & $-77\pm56^{+88}_{-43}$
                     & $1.0$ \\
Non-Res.             & $34.0\pm2.2^{+2.1}_{-1.8}$  
                     & $\delta^{\rm nr}_1=-11\pm5^{+3}_{-3}$
                     & $0$ (fixed)   & $-$ & $-$   \\
                     &               
                     & $\delta^{\rm nr}_2=185\pm20^{+62}_{-19}$  \\  \hline
$\chic K^\pm$        & $1.12\pm0.12^{+0.24}_{-0.08}$
                     & $-118\pm24^{+37}_{-38}$
                     & $0.15\pm0.35^{+0.08}_{-0.07}$
                     & $-77\pm94^{+154}_{-11}$
                     & $0.7$ \\
\hline \hline
  \end{tabular}
\end{table*}

\begin{figure}[b]
  \centering
  \includegraphics[width=0.48\textwidth]{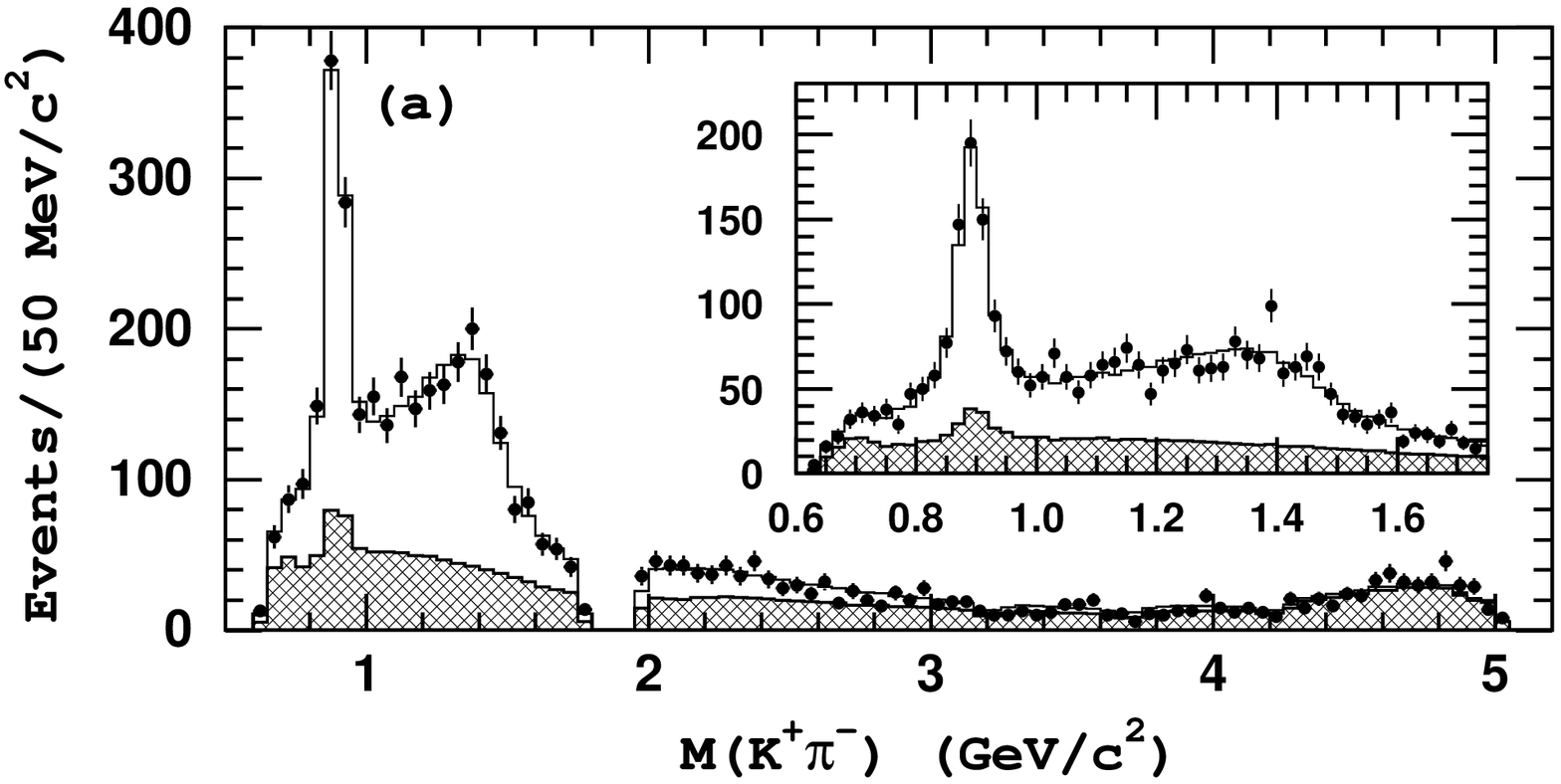}
  \includegraphics[width=0.48\textwidth]{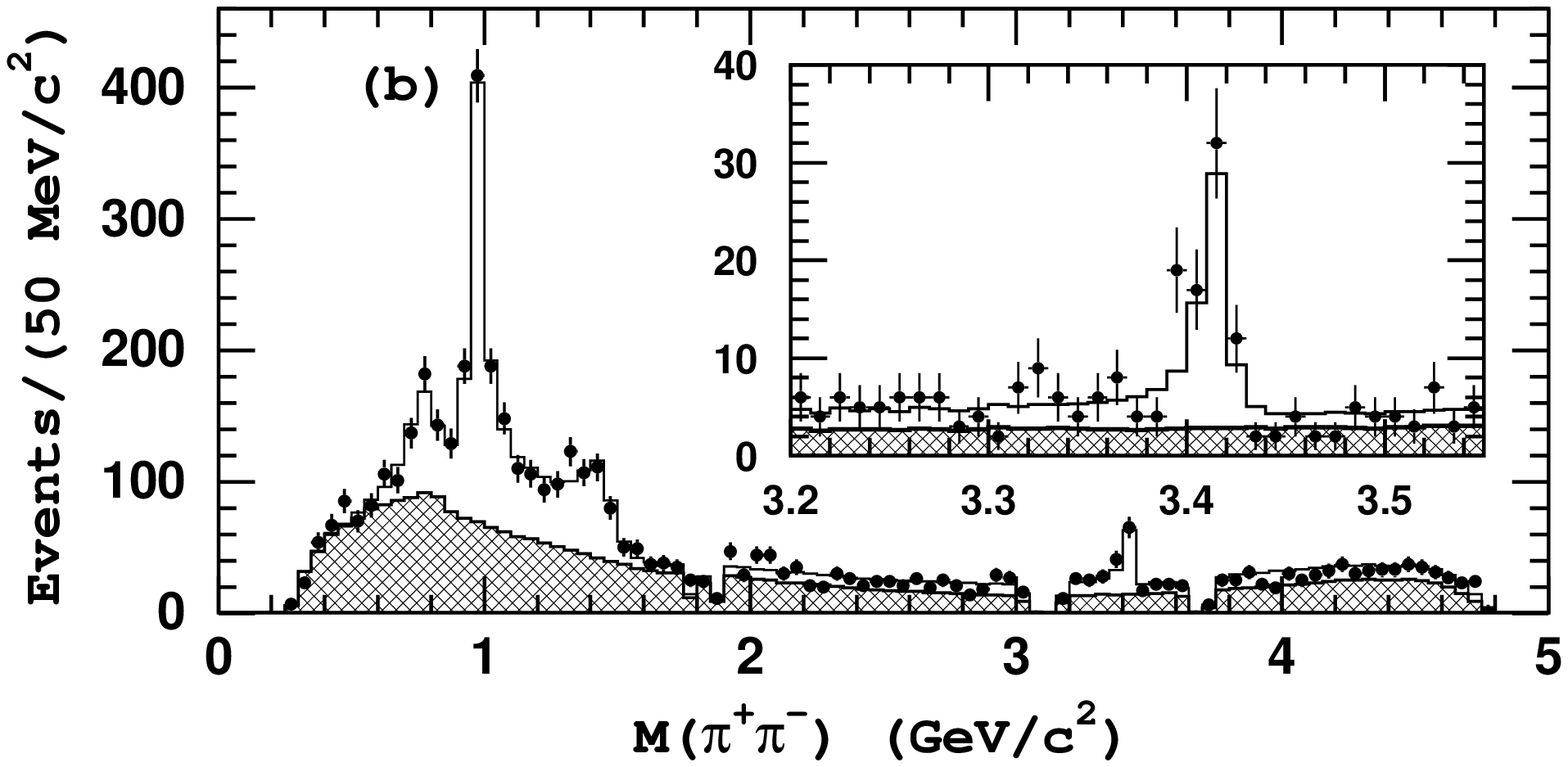}
  \caption{
           Results of the fit to $\kpp$ events in the $B$ signal region:
           (a) $M(\kcpi)$ distribution with $M(\pipi)>1.5$~GeV/$c^2$;
           (b) $M(\pipi)$ distribution with $M(\kcpi)>1.5$~GeV/$c^2$.
           Points with error bars are data, the open histogram
           is the fit result and hatched histogram is the background
           component. Inset in (a) shows the $K^*(892)-K_0^*(1430)$
           mass region in 20~MeV/$c^2$ bins. Inset in (b) shows the
           $\chic$ mass region in 10~MeV/$c^2$ bins.}
\label{fig:kpp-mod-d0}
\end{figure}

The analysis is performed by means of an unbinned maximum likelihood fit.
The distribution of background events is parametrized by an empirical function
with 11 parameters\,\cite{khh-dalitz-belle}.
As found in Ref.\,\cite{khh-dalitz-belle}, the three-body $B^+$$\to\,$$\kppp$
amplitude is well-described by a coherent sum of $K^*(892)^0\pi^+$,
$K^*_0(1430)^0\pi^+$, $\rho(770)^0K^+$, $f_0(980)K^+$, $f_X(1300)K^+$ and
$\chic K^+$ quasi-two-body channels and a non-resonant amplitude. The
$f_X(1300)K^+$ channel was introduced in order
to describe an excess of signal events at $M(\pipi)\simeq 1.3$~GeV/$c^2$
(see Fig.~\ref{fig:kpp-mod-d0}(b)). The best fit is achieved assuming
$f_X(1300)$ is a scalar state; the mass and width determined from the fit
(see below) are consistent with those for $f_0(1370)$\,\cite{PDG}.
Each quasi-two-body amplitude includes
a Breit-Wigner function, a $B$ decay form-factor parametrized in a
single-pole approximation, a Blatt-Weisskopf factor\,\cite{blatt-weisskopf}
for the intermediate resonance decay, and a function that describes angular
correlations between final state particles. This is multiplied by a factor
of $ae^{i\delta}$ that describes the relative magnitude and phase of the
contribution.
The non-resonant amplitude is parametrized by an empirical function
${\cal A}_{\rm nr}(\kppp)
      = a_1^{\rm nr}e^{-\alpha{s_{13}}}e^{i\delta^{\rm nr}_1} 
      + a_2^{\rm nr}e^{-\alpha{s_{23}}}e^{i\delta^{\rm nr}_2},$
where $\alpha$, $a_i^{\rm nr}$ and $\delta_i^{\rm nr}$ are fit parameters,
$s_{13}\equiv M^2(K^+\pi^-)$, and $s_{23}\equiv M^2(\pipi)$.
In this analysis we modify the model by changing the parameterization of the
$f_0(980)$ lineshape from a Breit-Wigner function to a Flatt\'e
parameterization\,\cite{Flatte} and by adding two more channels:
$\omega(782)K^+$ and $f_2(1270)K^+$. For $CP$ violation studies the amplitude
for each quasi-two-body channel is modified from $ae^{i\delta}$ to
$ae^{i\delta}(1\pm be^{i\varphi})$, where the plus (minus) sign corresponds
to the $B^+$ ($B^-$) decay. With such a parameterization the charge asymmetry,
$\ACP$, for a particular quasi-two-body $B$$\to$$f$ channel is given by
\begin{equation}
A_{CP}(f)
      = \frac{N^--N^+}{N^-+N^+}
      = -\frac{2b\cos\varphi}{1+b^2}.
\label{eq:acp-dcpv}
\end{equation}
It is worth noting that in this parametrization we assume zero relative phase
between $B^-$ and $B^+$ amplitudes.

\begin{figure}[b]
  \centering
\includegraphics[width=0.48\textwidth,height=80mm]{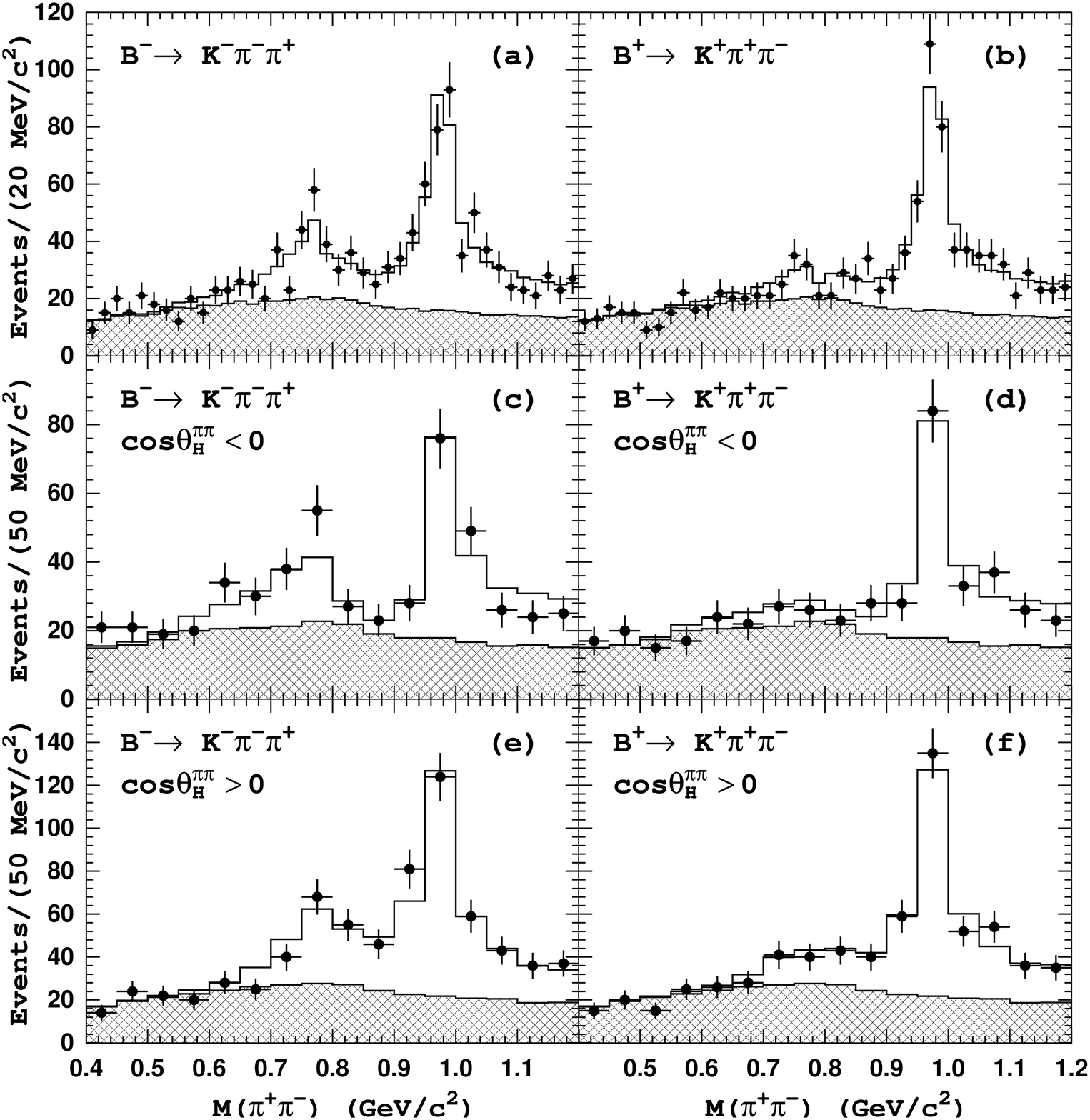}
  \caption{$\pipi$ mass spectra for $B^-$ (left column) and $B^+$ (right
           column) events for different helicity regions:
           (a,b) no helicity cuts;
           (c,d) $\cos\theta_H^{\pi\pi}<0$;
           (e,f) $\cos\theta_H^{\pi\pi}>0$;
           Points with error bars are data, the open histogram is the fit
           result and the hatched histogram is the
           background component.}
\label{fig:kpp-pp}
\end{figure}

First we fit the data with $b_i\equiv 0$ (no $CP$ violation) and determine the
parameters of the $f_X(1300)$ ($M=1.449\pm0.013(\rm stat.)$\,\,GeV/$c^2$,
$\Gamma=0.126\pm0.025(\rm stat.)$\,\,GeV/$c^2$),
$f_0(980)$ ($M=0.950\pm0.009(\rm stat.)$~GeV/$c^2$ and coupling constants
$g_{\pi\pi}=0.23\pm0.05(\rm stat.)$ and $g_{KK}=0.73\pm0.30(\rm stat.)$), and
the parameter of the non-resonant amplitude $\alpha=0.195\pm0.018(\rm stat.)$.
We then fix these six parameters and repeat the fit to data with $b$ and
$\varphi$ floating for all terms except $B^\pm$$\to\,$$\omega(782)K^\pm$ and
the non-resonant amplitudes. Possible effects of these assumptions were
studied, and are included in the final results as a part of the model
uncertainty. Projections of the fit are shown in Fig.\,\ref{fig:kpp-mod-d0},
and the results are summarized in Table\,\ref{tab:kpp-fit-res}. We find that
the statistical significance of the $B^\pm$$\to$$f_2(1270)K^\pm$ signal
exceeds $6\sigma$; this is the first observation of this decay mode. The
significance of the $B^\pm$$\to\,$$\omega(782)K^\pm$ signal is $2.1\sigma$.
The statistical significance of these signals (and the asymmetries quoted in
Table\,\ref{tab:kpp-fit-res}) is calculated as
$\sqrt{-2\ln({\cal L}_0/{\cal L}_{\rm max})}$, where ${\cal L}_{\rm max}$ and
${\cal L}_0$ denote the maximum likelihood with the nominal fit and with the
corresponding amplitude (or asymmetry) fixed at zero, respectively.
Note that the significance of the asymmetry is sensitive not only to
$b_i\cos\varphi_i$, but also to $b_i\sin\varphi_i$, whereas $\ACP$ is sensitive
only to the former (Eq.\,\ref{eq:acp-dcpv}). Therefore, the significance of the
asymmetry should be interpreted as having two degrees of freedom.
The only channel where the significance of the
asymmetry exceeds the $3\sigma$ level is $B^\pm$$\to$$\rho(770)^0K^\pm$.
Figures\,\ref{fig:kpp-pp}(a,b) show the $M(\pipi)$ distributions for the
$\rho(770)$$-$$f_0(980)$ mass region separately for $B^-$
and $B^+$ events. However, the interference term between $B$$\to$$\rho(770)K$
vector and $B$$\to$$f_0(980)K$ scalar amplitudes cancels out when making the
$M(\pipi)$ projection for the entire range of the helicity angle
$\theta_H^{\pi\pi}$ of the $\pipi$ system (the angle between the kaon and the
pion of the opposite charge in the $\pipi$ rest frame.)
Thus only the difference in relative fractions can be observed
from comparison of Figs.\,\ref{fig:kpp-pp}(a) and\,\ref{fig:kpp-pp}(b).
The effect is more apparent in $M(\pipi)$ spectra for the two helicity angle
regions $\cos\theta_H^{\pi\pi}$$<$$0$ and $\cos\theta_H^{\pi\pi}$$>$$0$ shown
in Figs.\,\ref{fig:kpp-pp}(c-f). Here the difference in the
interference terms for the $B^-$ and $B^+$ decay amplitudes (due to different
relative phases between the $B$$\to$$\rho(770) K$ and $B$$\to$$f_0(980)K$
amplitudes) can be distinguished as a difference in the shape of the
$M(\pipi)$ spectra for $B^-$ and $B^+$ decays. Results of the branching
fraction and $\ACP$ measurements are summarized in Table\,\ref{tab:results},
where for intermediate resonance fractions we use world average
values\,\cite{PDG}. The reconstruction efficiency is $22.4\pm0.2$\,\%,
determined from signal MC simulation in which events are generated according
to the matrix elements obtained from the best fit to data.

To assess how well any given fit represents the data, the Dalitz plot is
subdivided into non-equal bins requiring that the number of events in each
bin exceeds 25. A pseudo-$\chi^2$ variable for the multinomial distribution
is then calculated as
$ \chi^2 = -2\sum^{N_{\rm bins}}_{i=1}n_i\ln\left(\frac{p_i}{n_i}\right),$
where $n_i$ is the number of events observed in the $i$-th bin, and $p_i$ is
the number of predicted events from the fit. More details are given in
Ref.\,\cite{khh-dalitz-belle}. The $\chi^2/N_{\rm bins}$ value of the fit to
signal events is $182.5/141$ ($32$ fit parameters) and $127.6/120$ for
the fit to background events.

The following sources of systematic error are found to be dominant in the
determination of branching fractions:
charged track reconstruction (3\% in total);
particle identification efficiency (4.5\% in total);
requirements on event shape variables (2.5\%);
signal yield determination from the $\de$ fit (3.9\%);
model dependence (1\%);
number of produced $\bbbar$ pairs (1\%).
For the quasi-two-body channels additional sources 
are the uncertainty in parametrization of the background density function and
the uncertainty in secondary branching fractions (2\% for $f_2(1270)$,
11\% for $K^*_0(1430)$ and 10.8\% for $\chic$\,\cite{PDG}).
Note that in the asymmetry calculation most of these systematic
uncertainties cancel out. A few remaining sources are uncertainty due
to a possible asymmetry in background from charmless $B$ decays (0.6\%);
the possible bias due to intrinsic detector asymmetry (1.6\%);
$B^\pm$ signal yields determination (1.1\%).

\begin{table*}[!t]
  \caption{Summary of branching fraction results. The first quoted error is
           statistical, the second is systematic and the third is the model
           uncertainty. Note that $B^+\to\chic K^+$ contribution
           is not included in the three-body charmless branching fraction.}
  \medskip
  \label{tab:results}
\centering
  \begin{tabular}{lccr} \hline \hline
Mode & {\small $\BF(B^\pm\to Rh^\pm\to \kpp)\times10^{6}$} &
$\BF(B^\pm\to Rh^\pm)\times10^{6}$ & \multicolumn{1}{c}{$A_{CP}$ (\%)}
 \\ \hline \hline
 $\kpp$ Charmless       & $48.8\pm1.1\pm3.6$
                        &    $-$
                        & $+4.9\pm2.6\pm2.0$  \\
~$K^*(892)[K^\pm\pi^\mp]\pi^\pm$ 
                        & $6.45\pm0.43\pm0.48^{+0.25}_{-0.35}$
                        & $9.67\pm0.64\pm0.72^{+0.37}_{-0.52}$ 
                        & $-14.9\pm6.4\pm2.0^{+0.8}_{-0.8}$  \\
~$K^*_0(1430)[K^\pm\pi^\mp]\pi^\pm$
                        & $32.0\pm1.0\pm2.4^{+1.1}_{-1.9}$
                        & $51.6\pm1.7\pm6.8^{+1.8}_{-3.1}$
                        & $+7.6\pm3.8\pm2.0^{+2.0}_{-0.9}$   \\
~$\rho(770)^0[\pi^+\pi^-]K^\pm$
                        & $3.89\pm0.47\pm0.29^{+0.32}_{-0.29}$
                        & $3.89\pm0.47\pm0.29^{+0.32}_{-0.29}$
                        & $+30\pm11\pm2.0^{+11}_{-4}$     \\
~$f_0(980)[\pi^+\pi^-]K^\pm$
                        & $8.78\pm0.82\pm0.65^{+0.55}_{-1.64}$
                        & $-$                                 
                        & $-7.7\pm6.5\pm2.0^{+4.1}_{-1.6}$   \\
~$f_2(1270)[\pi^+\pi^-]K^\pm$
                        & $0.75\pm0.17\pm0.06^{+0.11}_{-0.18}$
                        & $1.33\pm0.30\pm0.11^{+0.20}_{-0.32}$
                        & $-59\pm22\pm2.0^{+3}_{-3}$   \\
~Non-resonant
                        & $-$
                        & $16.9\pm1.3\pm1.3^{+1.1}_{-0.9}$
                        & \multicolumn{1}{c}{$-$}     \\
\hline
 $\chic[\pi^+\pi^-]K^\pm$
                        & $0.56\pm0.06\pm0.04^{+0.12}_{-0.04}$
                        & $112\pm12\pm18^{+24}_{-8}$
                        & $-6.5\pm20\pm2.0^{+2.9}_{-1.4}$     \\
\hline \hline
  \end{tabular}
\end{table*}

To estimate model uncertainty in the branching fractions and
$\ACP$ for individual quasi-two-body channels, we vary the nominal model and
repeat the fit to data. The following variations are performed separately:
we add one additional channel which is either
$K^*(1410)^0\pi^+$, $K^*(1680)^0\pi^+$, or $K^*_2(1430)^0\pi^+$; remove
$\omega(782)K^+$ or $f_2(1270)K^+$ channel from the nominal model; fit the
data assuming $f_X(1300)$ is a vector ($\rho(1450)$) or excluding this
contribution; and use several alternative parameterizations of the non-resonant
amplitude\,\cite{khh-dalitz-belle,babar-kpp-dcpv}.
To cross check the asymmetry observed in $B^\pm$$\to$$\rho(770)^0K^\pm$,
we make an independent fit to $B^-$ and $B^+$ subsamples. We also confirm
the significance of the asymmetry observed in $B^\pm$$\to$$\rho(770)^0K^\pm$
channel with MC pseudo-experiments where events are distributed according to
the matrix element determined from the fit to data. All the cross-checks give
consistent results. Finally note that the second solution with a much smaller
fraction of the $K^*(1430)\pi$ signal as found in
Ref.\,\cite{khh-dalitz-belle} is confirmed in this analysis. However,
comparisons with results on elastic \mbox{$K$-$\pi$} scattering\,\cite{LASS}
and with some theoretical considerations\,\cite{chernyak} favor the solution
with a large $K^*(1430)\pi$ fraction. We find that values of $CP$ parameters
$b$ and $\varphi$ are almost solution-independent;
variation in their values is considered as a part of the model uncertainty.

In conclusion, we have performed an amplitude analysis of the three-body
charmless $\bckpp$ decay. The branching fractions for a number of
quasi-two-body channels have been measured; we report the first observation of
$B^+$$\to$$f_2(1270)K^+$, a tensor-pseudoscalar decay. We also perform a search
for direct $CP$ violation in quasi-two-body intermediate states and find
evidence for large direct $CP$ violation in the decay
$B^+$$\to$$\rho(770)^0K^+$. This is consistent with recent
results from BaBar\,\cite{babar-kpp-dcpv} and with some theoretical
predictions\,\cite{beneke-neubert}. The statistical significance of the
asymmetry is $3.9\sigma$ and varies from $3.7\sigma$ to $4\sigma$ depending
on the model used to fit the data. This is the first evidence for $CP$
violation in the decay of a charged meson. Statistical significance of the
$\ACP$ for the $B^\pm$$\to$$\rho(770)^0K^\pm$ channel is $3.0\sigma$.


We thank the KEKB group for the excellent operation of the
accelerator, the KEK cryogenics group for the efficient
operation of the solenoid, and the KEK computer group and
the NII for valuable computing and Super-SINET network
support. We acknowledge support from MEXT and JSPS (Japan);
ARC and DEST (Australia);\,NSFC (contract No.\,\,10175071,
China); DST (India); the BK21 program of MOEHRD and the CHEP
SRC program of KOSEF (Korea);
KBN (contract No.\,\,2P03B 01324,\,\,Poland);
MIST\,\,(Russia); MHEST (Slovenia);  SNSF (Switzerland);
NSC and MOE (Taiwan); and DOE (USA).




\end{document}